\documentclass[aps,pre,preprint,pdftex]{revtex4-2}

\usepackage{bm}
\usepackage{color}

\usepackage{amsmath}
\usepackage{graphicx}
\usepackage{hyperref}

\usepackage{ulem}

\usepackage{tabularx}
\newcolumntype{C}[1]{>{\hfil}m{#1}<{\hfil}}

\begin{document}

\title{
Fold analysis of crumpled sheet using micro computed tomography
} 

\author{Yumino Hayase}
\affiliation{Department of Physics, Kyushu University, Fukuoka
819-0395, Japan }

\author{Hitoshi Aonuma}
\affiliation{Research Institute for Electronic Science,
Hokkaido University, Sapporo 060-0812, Japan}

\author{Satoshi Takahara}
\affiliation{Graduate School of Information Science and Technology,
Hokkaido University, Sapporo 060-0812, Japan}

\author{Takahiro Sakaue}
\affiliation{Department of Physical Sciences, 
Aoyama Gakuin University,
5-10-1 Fuchinobe, Chuo-ku, Sagamihara, Kanagawa 252-5258, JAPAN}

\author{Shun'ichi Kaneko}
\affiliation{
Kazusa DNA Research Institute,
2-6-7 Kazusa-kamatari, Kisarazu, Chiba 292-0818, Japan
}

\author{Hiizu Nakanishi}
\affiliation{Department of Physics, Kyushu University, Fukuoka
819-0395, Japan}

\date{\today}

\begin{abstract}

Hand crumpled paper balls involve intricate structure with a network
  of creases and vertices, yet show simple scaling properties,
  which suggests self-similarity of the structure.  
  We investigate the internal structure of crumpled papers by 
the micro computed tomography (micro-CT)
  without destroying or unfolding them.
  From the reconstructed three dimensional data, we examine several
  power laws for the crumpled square sheets of paper of the sizes
  $L=50\sim 300$ mm, and obtain
the mass fractal dimension $D_M = 2.7\pm 0.1$ by the relation between
  the mass and the radius of gyration of the balls, and the fractal
  dimension $2.5\lesssim d_f \lesssim 2.8$ for the internal structure
  of each crumpled paper ball by the box counting method in the real
  space and the structure factors in the Fourier space;
  The data for the paper sheets are consistent with $D_M = d_f$,
suggesting that the self-similarity in the structure of each crumpled
  ball gives rise to the similarity among the balls with different
  sizes.
  We also examine the cellophane sheets and the aluminium foils of the
  size $L=200$ mm and obtain $2.6\lesssim d_f\lesssim 2.8$ for both of
  them.
The micro-CT also allows us to reconstruct 3-d structure of a line
drawn on the crumpled sheets of paper.  The Hurst exponent for
the root mean square displacement along the line is
estimated as $H\approx 0.9$ for the length scale shorter than the
scale of the radius of gyration, beyond which the line structure
becomes more random with $H\sim 0.5$.

\end{abstract}

\pacs{}

\maketitle

\section{Introduction}

Crumpling a sheet of paper is the easiest way to make effective shock
absorbing buffer.  A hand crumpled paper ball is very light with
typically more than 80\% of its volume being empty, still shows strong
resistance against compression.  These properties make it ideal spacer
for box packing.
Origin of these properties is the large stretching energy  in
comparison with the bending energy for a thin paper sheet. As a result,
upon crumpling a sheet, Gaussian curvature remains close to zero 
everywhere
except for at singular points of developable cone
structures\cite{Lobkovsky-1995,Cerda-1998,Cerda-1999,Witten-2007}.  This
imposes stringent constraint on the way how the paper sheet crumples,
thus produces strong resistance against compression even if 
much of the space is still empty\cite{Balankin-2007a,Lin-2008}.

In spite of their complex structure, the balls of crumpled sheet have
been known to show simple scaling
laws\cite{Kantor-1986,Kantor-1987,Gomes-1987}.  The radius of the ball
$R$ follows the scaling relation with the size of the original sheet
$L$ as
\begin{equation}
   R \sim L^\alpha,
  \label{R-L_scaling}
\end{equation}
and  with the force $F$ applied to make the crumpled ball as
\begin{equation}
  R \sim F^{-\delta}.
  \label{R-F_scaling}
\end{equation}
The exponents have been estimated as 
 $0.80\lesssim \alpha\lesssim 0.95$ 
for aluminium foil and paper, and $\delta \approx 0.2$ for aluminium foil
\cite{Kantor-1986,Kantor-1987,Gomes-1987,Lin-2008,Balankin-2007a,
Habibi-2017,Deboeuf-2013}.
The small value of $\delta$ corresponds to the fact that
the crumpled sheets resists strongly against compression.

These two power law relations suggest the fractal structure in the object,
and 
may be combined into the single
relation\cite{Vliegenthart-2006,Balankin-2007a}
\begin{equation}
  {R\over h} \propto \left({L\over h}\right)^{\alpha}
  \left({F\over Y h}\right)^{-\delta}, \label{R-LF_scaling} 
\end{equation} 
where $Y$ and $h$ are the 2-d Young modulus and the sheet thickness,
respectively.  The scaling relations (\ref{R-L_scaling}) and
(\ref{R-F_scaling}), however,  represent different aspects of the
structure, namely, the former implies the scaling in the structure
among the crumpled balls with different sizes, and the latter suggests
self-similarity of internal structure of each crumpled
ball\cite{Balankin-2007b}\footnote{
It should be noted that the definition of the applied force $F$ and
the radius of crumpled ball $R$ depend on an experimental protocol.
Due to the plasticity and the relaxation involved during and after
crumpling, there is certain subtleties in 
what these values really mean,
but we will not go into this problem, simply
assuming that such ambiguity does not affect the
scaling relations we will study in the present work%
}.

The internal structure of the crumpled paper ball has been studied by
examining the crease networks on an unfolded sheets\cite{Blair-2005,
Balankin-2006,Andresen-2007,Balankin-2013a}, the cross sections
obtained by cutting the balls in
half\cite{Balankin-2010,Deboeuf-2013,Balankin-2013a}, or the sequences
of holes made by a needle piercing through the
balls\cite{Balankin-2007b} as well as numerical
simulations\cite{Vliegenthart-2006,Tallinen-2008,Tallinen-2009,Liou-2014}.
These are indirect way of observing the internal structure.  

The X-ray micro computed tomography (micro-CT) is computer
tomography with the high resolution of the order of 100 $\mu$m, and
makes it possible to study the internal structure of crumpled sheets
without either unfolding or destroying them.  It has been used to
study the crumpled balls of aluminium foil
to examine the density distribution, the curvatures of the sheets, and
the fractal dimensions of the
structure\cite{Cambou-2011,Lin-2009-PRL,Lin-2009-PRE}.  In this paper,
we use micro-CT to examine the internal structure of crumpled sheets
of paper, cellophane, and aluminium foils to determine
scaling properties of the structure.
Experimental procedure is described in Sec.II, the scaling analysis
for the observed quantities is provided in Sec.III, experimental
results are presented to obtain several exponents in Sec. IV, and the
discussions are given in Sec. V.

\section{Experimental procedure}\label{ExpProc}

\begin{table}
  \begin{tabular}{C{8em}| C{10em}C{10em}C{10em}}
& Tracing paper & Cellophane & Aluminium foil \\
     \hline\hline
Manufacturer & Kokuyo Co., Ltd. & Toyo Co.  & Shimojima Co., Ltd.  \\ \hline
Thickness [mm] 
  & 0.0385$\pm$ 0.0008 & 0.0203$\pm$0.0002 & 0.0115$\pm$0.0004 \\ \hline
Density [$\rm g/m^2$] 
    & 42.7$\pm$0.5 & 30.0$\pm$0.7 & 27.4$\pm$0.6 \\
    \hline
  \end{tabular}
  \caption{Sample specifications.}
  \label{table-I}
\end{table}

Square sheets of tracing paper with the side length $L=$ 50, 100, 200,
and 300 mm are hand crumpled without any specific protocol.
Before being scanned, the crumpled balls are 
left for 7 days under the environment with the temperature
$25\pm2^\circ$C and the humidity $25\pm5$\% to allow them to settle,
in order to avoid structure relaxation during about 30 minutes of the
scanning time\footnote{
See Supplemental Material at [url] for the relaxation effects during
the 7-day waiting time in the sample preparation.
}.
The X-ray micro-CT 
(inspeXio SMX-100CT, Shimadzu Corporation, Kyoto, Japan) 
is operated at 40 kV with 100 $\mu$A,
and
%
scans a cubic region of
the space with the linear size around $4$ cm at a resolution
50$\sim$150 $\mu{\rm m}$ with around $400^3$ voxels.  
We also scan hand crumpled cellophane sheets and aluminium foils of
the size $L=200$ mm 
following the same procedure described above except that, in the case
of aluminium foil, the samples are scanned on the day when they are
crumpled.
The thicknesses and densities of the samples are
listed in Table \ref{table-I}.

\begin{figure}
  \centering\includegraphics[width=8cm]{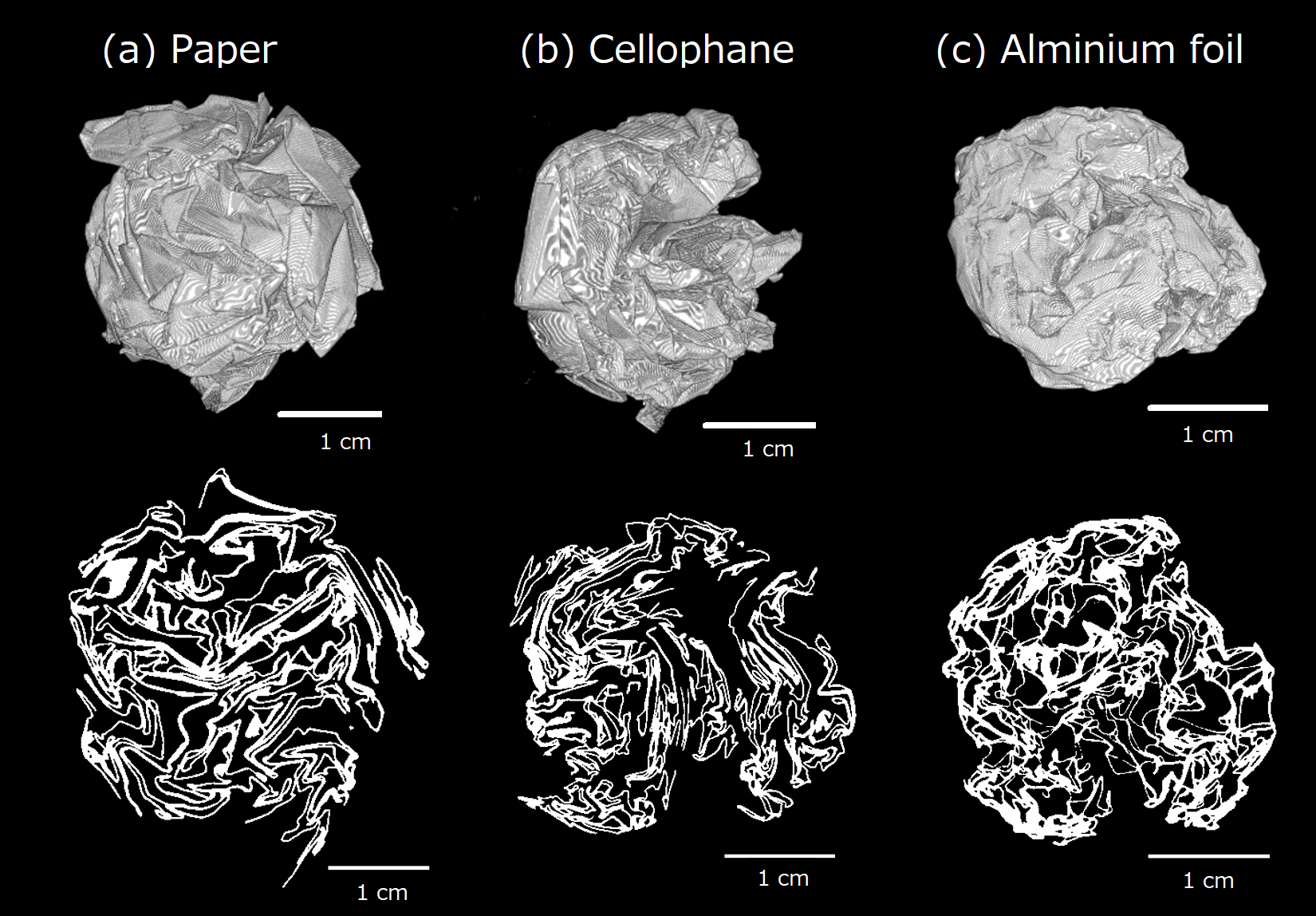}
  \caption{CT images of crumpled paper (a), cellophane sheet (b), 
  and aluminium foil (c).
  The upper images are reconstructed 3-d structures and the lower
  images are their cross sections.}
  \label{CT_images}
\end{figure}

The micro-CT produces grey scale data for each slice of cross section;
the black and white binary data are generated by setting appropriate
threshold, and then the 3-d structures are reconstructed.
Figure \ref{CT_images} shows examples of
the reconstructed 3-d structures from the CT images
and the cross sections for 
crumpled paper, cellophane sheet, and aluminium foil.

\begin{figure}
  \centering\includegraphics[width=8cm]{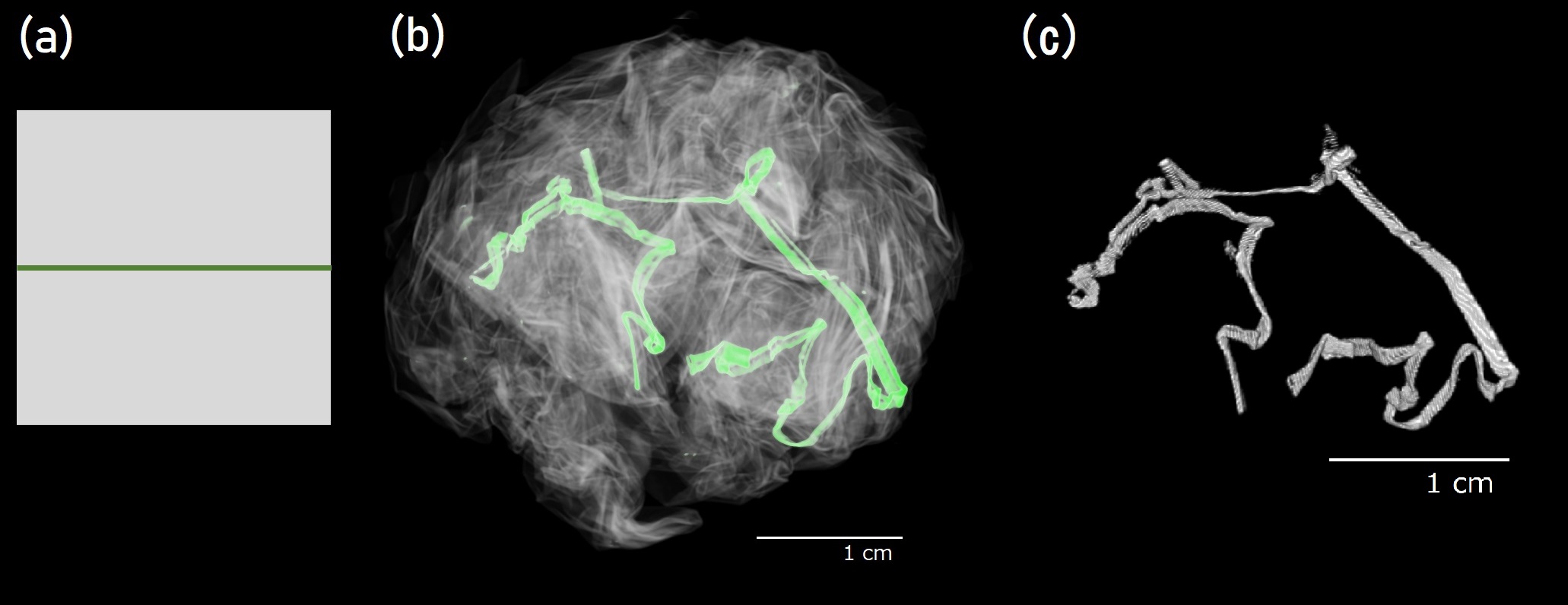}
  \caption{A line on a crumpled paper of the size $L=200$ mm.
(a) schematic illustration of a line drawn on a paper with the ink
  that contains tungsten,
(b) an example of three dimensional image of a crumpled paper with the line,
  and (c) the extracted 3-d image of the line.
}
  \label{Line-on-paper}
\end{figure}

We also reconstruct the 3-d configurations of a line drawn on a
crumpled paper (Fig.\ref{Line-on-paper}).  A straight line is drawn
using the ink that contains tungsten (Macky gold, Zebra,
Tokyo, Japan), which absorbs X-rays more efficiently than the paper.
Thus the line positions can be extracted from the CT images by setting
a higher threshold than that for the paper structure.


\section{Observed Quantities and Scaling Analysis}\label{sec-3}

Before we show our experimental results, we give some scaling analysis
for the physical quantities we observe.

\paragraph{Three dimensional structure of crumpled sheets:}

The mass fractal dimension $D_M$ is defined by
the scaling relation between the mass of the paper $M$ and
the radius of the crumpled paper ball $R$ as 
\begin{equation}
 M \sim R^{D_M}.
  \label{mass_fractal_dim}
\end{equation}
Since the mass of the paper $M$ is  proportional to its area,
$ M \propto L^2 $, 
the mass fractal dimension $D_M$ is related to the exponent $\alpha$ in
Eq.(\ref{R-L_scaling}) as
\begin{equation}
  D_M= 2/\alpha .
  \label{D_M=2/alpha}
\end{equation}
%
On the other hand,
the fractal dimension $d_f$ of the internal structure
of each crumpled ball is measured by the box counting method,
using the relation
\begin{equation}
  N_{\rm box} \sim l^{-d_f},
  \label{BoxCounting-d_f}
\end{equation}
where $N_{\rm box}$ is the number of occupied boxes with the linear
size $l$.

These two exponents correspond to two distinct features, i.e.
the fractal dimension $d_f$ defined by Eq.(\ref{BoxCounting-d_f}) describes
the self-similarity of each crumpled paper ball structure while
the mass fractal dimension $D_M$ represents the similarity among
crumpled paper balls with different size $L$.
However, the self-similar structure of each
crumpled paper ball suggests the self-similarity among those of
different sizes, and Eq.(\ref{BoxCounting-d_f}) could be extended as
\begin{equation}
  N_{\rm box}\sim {M\over h^2 \sigma} \left( {l\over h} \right)^{-d_f},
  \label{BoxCounting-d_f-2}
\end{equation}
where $h$ is  the paper thickness and
$\sigma$ is the area density of the sheet.
This leads to $d_f=D_M$ if $R$ is identified as the box size $l$ that
corresponds to $N_{\rm box}=1$, i.e.
\begin{equation}
  1 \sim {M\over h^2 \sigma} \left( {R\over h} \right)^{-d_f}.
\end{equation}


The structure factor $S(\bm q)$ is the Fourier transform of the
density correlation function $g(\bm r)$,
\begin{equation}
  S(\bm q) = 
   \iiint e^{-i\bm q\cdot\bm r} g(\bm r) d\bm r ,
\end{equation}
where the density correlation is defined by
\begin{equation}
  g(\bm r) = \iiint \big<\rho(\bm r'+\bm r)\rho(\bm r')\big> d\bm r'
\end{equation}
in terms of the density distribution $\rho(\bm r)$.
Here, $\left<\cdots\right>$ means the ensemble average.
If the structures of the crumpled paper ball are self-similar 
with the fractal dimension $d_f$,
then
the density correlation should be of the scaling form,
\begin{equation}
  g(r) \sim r^{d_f-3},
  \label{scaling_g}
\end{equation}
then the structure factor is expected to be of the scaling form
\begin{equation}
  S(q) \sim q^{-d_f}.
  \label{Sq-d_f}
\end{equation}
%
%

\paragraph{Cross section of a crumpled sheet:}
A 3-d CT image consists of  hundreds of two dimensional slices of
the density distribution in the cross sections.
We analyze the structure of the cross section which contains the
center of mass of the crumpled sheet.
Suppose that we take the center of mass of the crumpled sheet  as
the origin of the co-ordinate, and
consider the cross section by the $z=0$ plane.
Let $\sigma_{\rm cs}(\bm r_\perp)$ denote the 2-d density distribution
on the cross section as a function of a position $\bm r_\perp=(x,y)$ on the
$z=0$ plane,
\begin{equation}
\sigma_{\rm cs}(\bm r_\perp) = \int_{-h/2}^{h/2}\rho(\bm r_\perp,z) dz.
\end{equation}
The density correlation on the cross
section $g_{\rm cs}(\bm r_\perp)$ is defined by
\begin{equation}
  g_{\rm cs}(\bm r_\perp)
  = \iint\big<\sigma_{\rm cs}(\bm r'_\perp+\bm r_\perp)
                 \sigma_{\rm cs}(\bm r'_\perp)\big> d\bm r'_\perp,
\end{equation}
and the structure factor of the cross section $S_{\rm cs}(\bm q_\perp)$
is given by the 2-d Fourier transform
\begin{equation}
  S_{\rm cs}(\bm q_\perp) = \iint e^{-i\bm q_\perp\cdot\bm r_\perp}
  g_{\rm cs}(\bm r_\perp) d\bm r_\perp,
\end{equation}
where $\bm q_\perp$ is the wave vector within the cross section plane.
If we assume the same form as Eq.(\ref{scaling_g}) for $g_{\rm cs}$ as
\begin{equation}
  g_{\rm cs}(r_\perp) \sim r_\perp^{d_f-3},
\end{equation}
then the structure factor would behaves as
\begin{equation}
  S_{\rm cs}(q_\perp) \sim q_\perp^{-(d_f-1)},
  \label{S_cs(q)}
\end{equation}
which simply shows that the fractal dimension for the cross section is $d_f
-1$.

\paragraph{Straight line drawn on a crumpled paper:}
The CT technique allows us to study the structure of a line drawn on a
crumpled paper.  A straight line on a flat paper is deformed into a
random structure as the paper is crumpled.  The configuration of the
crumpled line can be represented by the function
\begin{equation}
  \bm r_{\rm line}(s) ; \qquad 0\le s \le L,
\end{equation}
where $s$ is the distance along the line from one of the end.  The
root mean square (RMS) distance $R_{\rm line}(s)$
from one of the end points is defined by
\begin{equation}
  R_{\rm line}(s) \equiv \sqrt{\left< \big(\bm r_{\rm line}(s)-\bm r_{\rm
  line}(0)\big)^2\right>}.
  \label{R_line}
\end{equation}
If this shows the power law behavior
with the Hurst exponent $H$, 
\begin{equation}
  R_{\rm line}(s) \sim s^H,
  \label{line-Hurst}
\end{equation}
then the scaling argument based on the self-similarity assumption leads to
the scaling law
\begin{equation}
  g_{\rm line}(r) \sim r^{-3+1/H}
\end{equation}
for the correlation of the line in 3-d space,
and the scaling law
\begin{equation}
  S_{\rm line}(q) \sim q^{-1/H}
  \label{S_q_line}
\end{equation}
for the 3-d structure factor of the line.

The size of the crumpled line $R_{\rm line}$ is expected to scale with
the size of the paper $L$ as
  \begin{equation}
    R_{\rm line} \sim L^{\alpha_{\rm line}}.
  \end{equation}
If the size of the line $R_{\rm line}$ should be of the same order
with the size of the crumpled ball $R$, and also with
RMS of the end-to-end distance
of the line $R_{\rm line}(L)$ of Eq.(\ref{line-Hurst}),
then 
the Hurst exponent should be related to the mass fractal dimension as
\begin{equation}
  H = \alpha_{\rm line} = \alpha = 2/D_M.
  \label{H=2/D_M}
\end{equation}

\section{Experimental Results}

%
We analyze the 3-d structures of the crumpled
sheet balls, the cross sections of the balls, and the lines drawn on the
paper.


\paragraph{Three dimensional structure of crumpled sheets:}

\begin{figure}
  \centering{\includegraphics[width=10cm]
{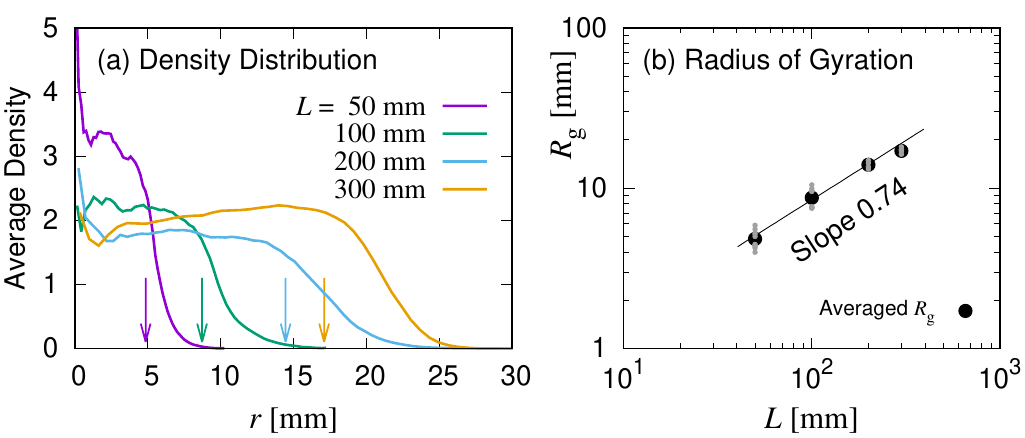}
  }
\caption{Density distributions (a) and radii of gyration (b) 
  for crumpled paper sheets of the size $L=$ 50, 100, 200, and
  300 mm.  
  (a) Each curve  represents average over around 10 samples and the
  arrows show the average values of the radii of gyration.  
  The density distributions are normalized so that the total mass
  should be proportional to $L^2$.
  (b) 
  The grey circles and the black
  circles represent the radius of gyration for each sample and
  average values over them, respectively.
  } 
\label{density-dist}
\end{figure}

Figure \ref{density-dist}(a) shows the averaged density distribution
of crumpled paper as a function of  the distance from the center of
mass $r$.  Each line represents the average distribution over the
direction and around 10 samples, and the arrows show the average
values of the radii of gyration for the corresponding sizes of the
paper.
One can see that the averaged density distribution inside the ball is
roughly uniform, but it is a slightly decreasing function of $r$ for
the crumpled balls of the smaller sheets $L=50$ and 100 mm, 
almost constant in the range of
$r\lesssim 12$ mm for that of $L=200$ mm, and slightly increasing in
$r\lesssim 16$ mm for that of $L=300$ mm.  
It is not clear how this tendency extends to larger sheets.
In Fig.\ref{density-dist}(b), the radii of gyration $R_g$ are plotted
against the paper size $L$ in the logarithmic scale. 
The data range is less than one decade and not enough to give
a precise value of the exponent, but
the plots are consistent with the power law behavior
\begin{equation}
 R_g \sim L^\alpha 
  \label{R_g-L-scaling}
\end{equation}
with the exponent $\alpha\approx 0.74$.
This gives $D_M=2/\alpha \approx 2.7$ from Eq.(\ref{D_M=2/alpha}).

\begin{figure}
  \centering{
\includegraphics[width=10cm]
{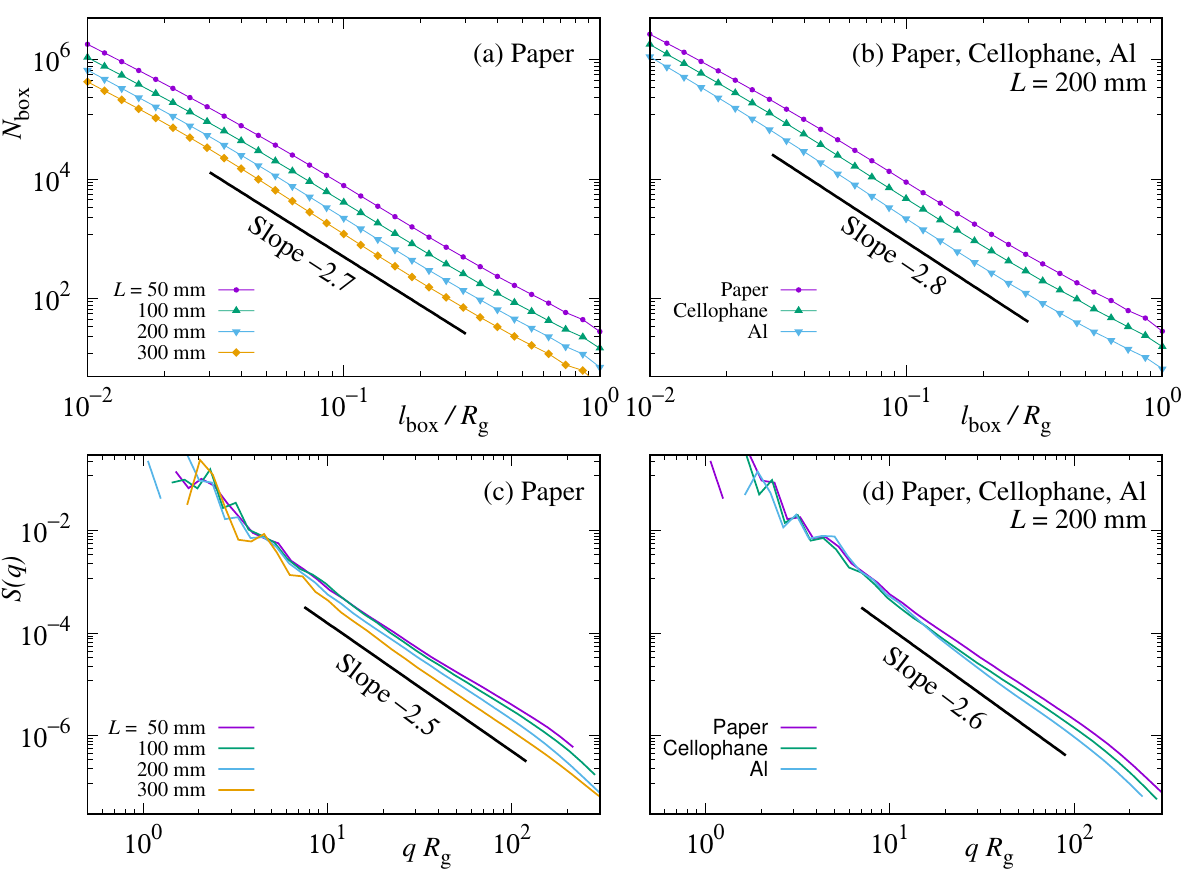}
}
  \caption{
 The box counting data and the averaged structure factors $S(q)$
 for the 3-d structures of crumpled paper, cellophane, and aluminium foil.
 The box size $l_{\rm box}$  and the wave number $q$ are scaled by the
 radii of gyration of each structure.  $N_{\rm box}$ for different
 size $L$ (a) and different materials (b) are shifted by the factor 2
 to avoid overlapping.}
  \label{Total-Structure}
\end{figure}

We estimate the fractal dimensions $d_f$  by the box
counting method;  the number of occupied boxes $N_{\rm box}$ are
plotted against the linear size $l_{\rm box}$ of the box divided by
$R_g$ in the logarithmic scale in Fig.\ref{Total-Structure}(a) for
paper and (b) for cellophane and aluminium foil.  Each data
point is an average of about 10 samples. 
The data for different size $L$ or different materials are shifted
vertically by multiplying by the factor 2 to avoid overlapping of the
plots.  The estimated $d_f$ for the crumpled paper is 2.7 from
Fig.\ref{Total-Structure}(a). 
This is consistent with $D_M$ estimated by $R_g$, and suggests
the self-similarity in the 3-d structure of the crumpled paper
ball as we have discussed.
The fractal dimensions $d_f$ 
for both the cellophane sheet and aluminium foil are estimated as 2.8 from
Fig.\ref{Total-Structure}(b).

The structure factors $S(q)$ for the reconstructed 3-d structures
from the CT data are plotted in the logarithmic scale
for the paper of sizes $L=50\sim 300$ mm in
Fig.\ref{Total-Structure}(c), and for the paper sheets, the cellophane
sheets, and the aluminium
foils of the size $L=200$ mm in Fig.\ref{Total-Structure}(d).
Each data point represents averaged value of about 10 samples.
These structure factors $S(q)$ show the power law behavior
\begin{equation}
  S(q) \sim q^{-\beta}.
  \label{S_q}
\end{equation}
The apparent value of the exponent $\beta$ for the crumpled paper in
Fig.\ref{Total-Structure}(c) increases with the paper size $L$ and the
fitted value for the largest paper size $L=300$ mm is $\beta\approx
2.5$.  
The exponents $\beta$ for the cellophane sheets and the aluminium foils
are estimated
from the plots for $L=200$ mm in Fig.\ref{Total-Structure}(d) 
as $\beta\approx 2.6$.
These values of $\beta$ estimated by $S(q)$ are slightly smaller than
those of $d_f$ estimated by the box counting method although they
should coincide as
\begin{equation}
  \beta = d_f
  \label{beta=d_f}
\end{equation}
from Eq.(\ref{Sq-d_f}) if the self-similarity holds.

\paragraph{Cross section structure of crumpled sheets:}

\begin{figure}
  \centering{\includegraphics[width=10cm]
{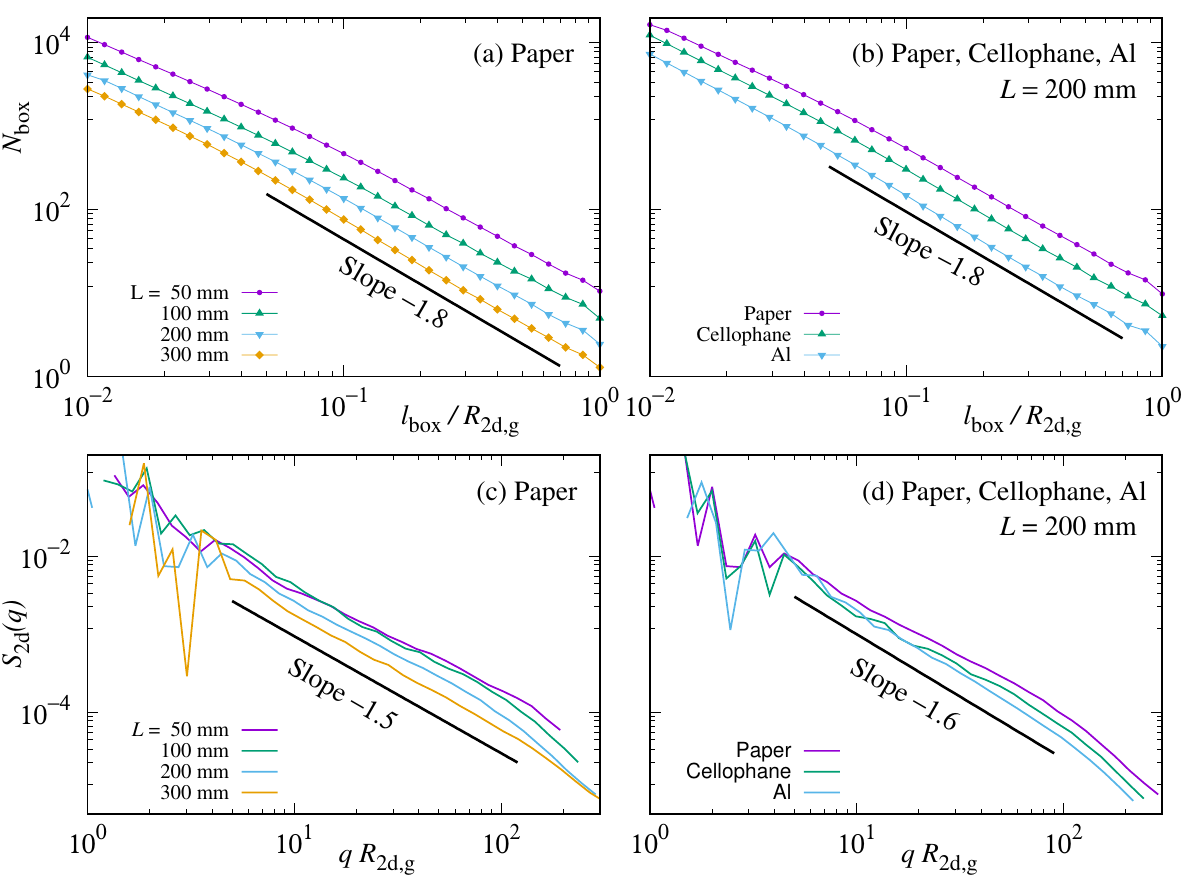}
  }

  \caption{
    The box counting numbers $N_{\rm box}$ 
    and the averaged structure factors $S_{2d}(q)$ 
    for the cross sections of crumpled paper sheets,
  cellophane sheets, and aluminium foils.  
  The box size $l_{\rm box}$ and the wave number $q$ 
  are scaled by the radii of gyration of each structure.
  $N_{\rm box}$ for different size $L$ (a) and different materials
  (b) are shifted by the factor 2 to avoid overlapping.}
  \label{Cross-Section}
\end{figure}

The structure of the cross section of the crumpled sheet is examined
in the same way.  Figure \ref{Cross-Section} shows
the box counting data and the 2-d structure factor 
for the cross section of the crumpled
sheets; The horizontal axes are scaled by the 2-d radius of gyration

The fractal dimensions $d_{f, \rm cs}$ for the cross sections are
estimated  by the box counting method, and we obtain
$  d_{f, \rm cs} \approx 1.8$
for the paper  of the size $L=50 \sim 300$ mm, and
$  d_{f, \rm cs} \approx 1.8 $
for cellophane and aluminium of the size $L=200$ mm.
The 2-d structure factor also shows the power law behavior
\begin{equation}
  S_{2d}(q_\perp)\sim q^{-\beta_{\rm cs}}
\end{equation}
with the exponent
 $ \beta_{\rm cs} \approx 1.5 $ for the paper
and
$\beta_{\rm cs} \approx 1.6$ for the cellophane and the aluminium.
Since the self-similarity leads to the relation
\begin{equation}
  \beta_{\rm cs} = d_{f,\rm cs} = d_f-1
  \label{beta_cs=d_f-1}
\end{equation}
from Eq.(\ref{S_cs(q)}), the obtained values for $\beta_{\rm cs}$ are
consistent with the corresponding exponents for the 3-d structure.

\paragraph{Structure of a line on a crumpled paper:}

\begin{figure}
  \centering{\includegraphics[width=10cm]
  {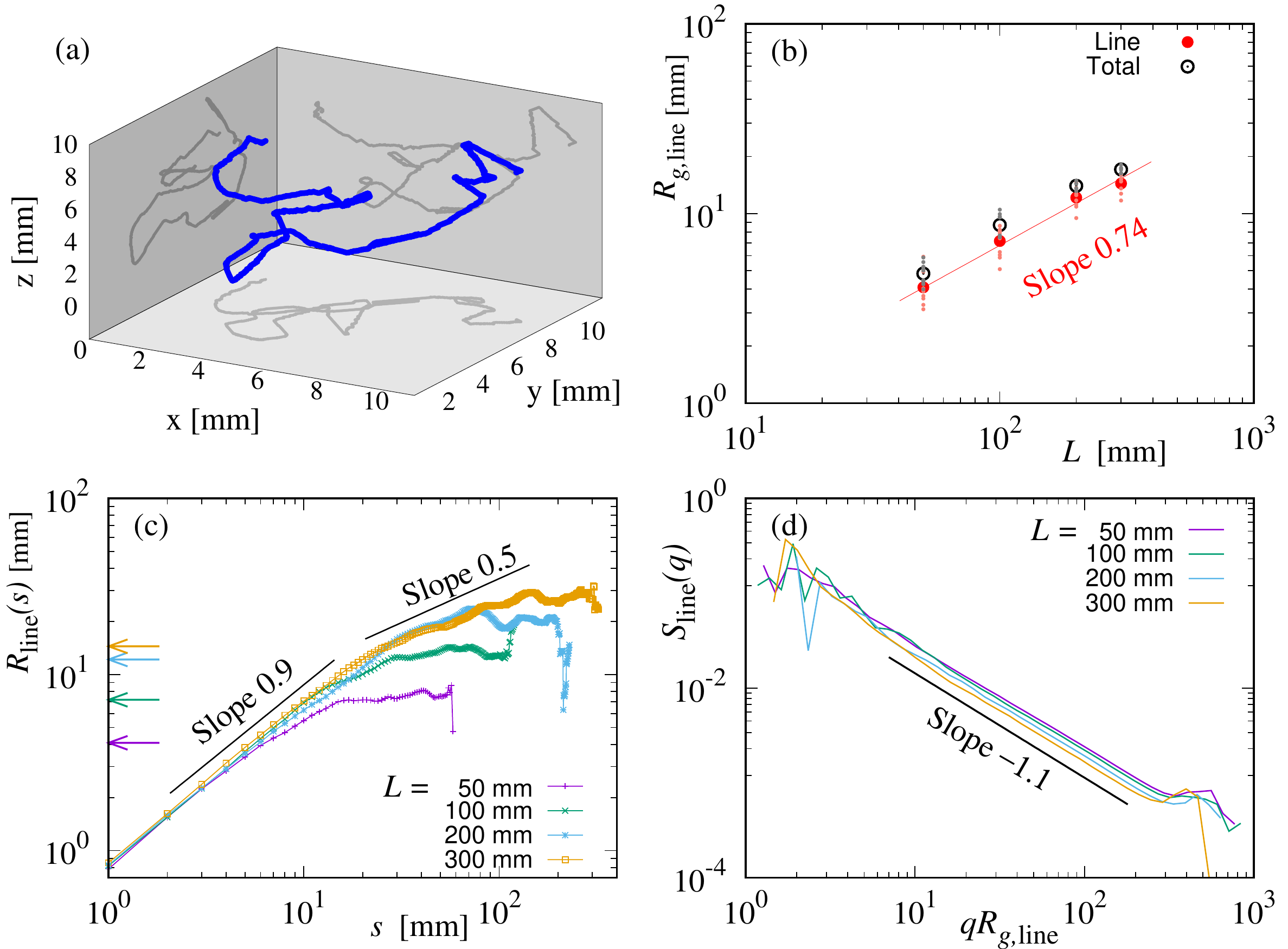}}

  \caption{The structure of a line drawn on a paper.
  (a) An example of a three dimensional configuration of a line
  on a crumpled paper of the size $L=50$ mm.
  (b) Radius of gyration v.s. the paper size $L$.
  (c) Averaged root mean square displacement at the position $s$ from
  one of the ends along the line. The arrows indicate the $R_{\rm
  line,g}$ for each size of the paper.
  (d) Averaged structure factors $S(q)$. The wave number $q$ is scaled
  by $R_{g, \rm line}$ for each size of the paper.
  }
  \label{Line-str}
\end{figure}

Fig.\ref{Line-str}(a) shows a configuration of a line one a crumpled
paper.
The radii of gyration $R_{g, \rm line}$ for the lines are plotted as a
function of the paper size $L$ in Fig.\ref{Line-str}(b) along with
$R_{g}$ for the whole structure as have been plotted in  
Fig.\ref{density-dist}(b);  
the values of $R_{g, \rm line}$ are somewhat smaller than those of
$R_g$, but the plot is consistent with the power law behavior with the
same exponents as it should be, i.e.
\begin{equation}
  R_{g, \rm line}\sim L^{\alpha_{\rm line}}.
\end{equation}
with $\alpha_{\rm line}\approx 0.74$.
In Fig.\ref{Line-str}(c),
RMS distance $R_{\rm line}(s)$ defined by Eq.(\ref{R_line})
is plotted as a function of $s$ in the logarithmic scale to estimate
the Hurst exponent $H$.
The plot shows roughly the power law behavior
\begin{equation}
  R_{\rm line}(s) \sim 
    s^H 
    \label{R_line-H}
\end{equation}
with  $H\approx 0.9$ for the range $R_{\rm line}(s)\lesssim R_{g,\rm
line}$, beyond which it saturates and  seems to follow the power law
with a smaller exponent $H\approx 0.5$, but the range is too small to
determine its behavior with confidence.
%
Finally, Fig.\ref{Line-str}(d) shows that the 3-d structure factor for
the line $S_{\rm line}(q)$ behaves as
\begin{equation}
  S_{\rm line}(q) \sim q^{-\beta_{\rm line}} 
  \quad \mbox{with }\beta_{\rm line}\approx 1.1.
  \label{S_q-line-2}
\end{equation}
The results by Eqs.(\ref{R_line-H}) and (\ref{S_q-line-2}) are
consistent with the relation (\ref{S_q_line}), but
not with the relation (\ref{H=2/D_M}).

\begin{table}
  \begin{tabular}{l|C{3em}C{3em}C{3em}|C{3em}C{3em}|C{3em}C{3em}C{3em}|}
    &\multicolumn{3}{c|}{3-d structure}
    &\multicolumn{2}{c|}{cross section}
    &\multicolumn{3}{c|}{line}\\
    %
    & $\alpha$ & $\beta$ & $d_f$
    & $\beta_{\rm cs}$ & $d_{f,\rm cs}$ 
    & $\alpha_{\rm line}$ & $H$ & $\beta_{\rm line}$ \\
    \hline \hline
    Paper ($L=50\sim 300$ mm)
    & 0.74 & 2.5 & 2.7 &   1.5 & 1.8 
    & 0.74 & 0.9 & 1.1 \\
    Cellophane ($L=200$ mm)
    & $-$ & 2.6 & 2.8 &  1.6 & 1.8 
    & $-$ & $-$ & $-$   \\
    Aluminium foil ($L=200$ mm)
    & $-$ & 2.6 & 2.8 &  1.6 & 1.8 
    & $-$ & $-$ & $-$   \\
    \hline
  \end{tabular}
  \caption{Obtained exponents for the crumpled paper, cellophane, and
  aluminium foils.}
  \label{table-II}
\end{table}

\section{Discussions}

We have estimated several exponents which describe the scaling
behaviors of the structure of crumpled sheets; The results are
tabulated in Table \ref{table-II}.  Considering the range of the data
points and the data fluctuations, error bars for each exponent would
be around $\pm 0.1$.

As is described in Sec.\ref{ExpProc}, the paper and cellophane samples
are scanned 7 days after they are crumpled.  This is to avoid the
structural relaxation immediately after crumpling and to obtain steady
values of measurement.  One might wonder if the 7-day waiting time is
enough to obtain steady values, especially when logarithmically slow
relaxation has been observed up to three weeks in similar systems of
crumpled thin sheets\cite{Matan-2002, Balankin-2006, Lahini-2017}.
In these works, the slow relaxation is observed
in the compaction height under a constant force\cite{Matan-2002}, the
diameter of crumpled ball after the folding force is
withdrawn\cite{Balankin-2006}, and the stress under a constant
compression\cite{Lahini-2017}.
We have checked if a similar slow relaxation shows any significant effect
on the quantities we measure in the present work, but
it turns out that the relaxation effects are virtually
invisible in the scaling exponents of the structure beyond one hour
after crumpling (See Supplemental Material).

As we have discussed in Sec.\ref{sec-3}, the exponents in Table
\ref{table-II} are related to the two basic exponents: the mass
fractal dimension $D_M$ and the fractal dimension $d_f$.  The former
represents the scaling behavior among the crumpled balls of the
different sheet sizes while the latter describes the self-similarity
of the structure of each crumpled sheet.
For the paper sheet of the size $L=50\sim 300$ mm, the mass fractal
dimension estimated from $\alpha$ by Eq.(\ref{D_M=2/alpha}) is
$D_M\approx 2.7$.  The fractal dimension for the internal structure
$d_f$ are estimated from the box counting and the structure factor;
The structure factor tends to give smaller values for the fractal
dimension as has been found also in \cite{Lin-2009-PRE}, but
overall data suggest that the fractal dimension is in the range
$2.5 \lesssim d_f \lesssim 2.8$.
These estimates are consistent with
\begin{equation}
 D_M = d_f 
\end{equation}
within the accuracy of our estimate, suggesting that 
the self-similarity in the structure of each crumpled ball
gives rise to the similarity among the balls with different
sizes.
%
For the cellophane and the aluminium with the size $L=$ 200 mm, 
the estimated fractal dimension for the structure 
$2.6\lesssim d_f\lesssim 2.8$, 
which are slightly larger than that for the paper;
We do not have data to estimate $D_M$ for the cellophane and the
aluminium.

These values for $D_M$ and $d_f$ are somewhat larger than the values
obtained for $D_M$ in previous works: 2.51 \cite{Gomes-1987} and
2.1$\sim$2.5 \cite{Balankin-2007b} for paper, and
2.5 \cite{Kantor-1986,Kantor-1987} and 2.3 \cite{Balankin-2007a} for
aluminium foil.
In these estimates, the external diameter is used for the size of the
ball $R$ while in the present work the radius of gyration $R_g$
calculated from the density distribution is used.  This may lead to
some difference in estimating $D_M$ especially when the data range is
not large enough although they should give the same exponent in the
limit of the infinite data range.  
Note that the fractal
dimensions obtained in the present work for cellophane and aluminium
foil are for the internal structure of each crumpled ball.

The unique piece of information that the micro-CT can provide is the
structure of a line drawn of the crumpled paper sheets.  The estimated
value of the Hurst exponent $H\approx 0.9$ for the short length scale
suggests that the line configuration is quite ballistic for the length
scale up to $R_g$, but it eventually
approaches the random walk for the longer scale.
The value of the Hurst exponent $H\approx 0.9$ is consistent with the
relation (\ref{S_q_line}) to the exponent for the structure factor
$\beta_{\rm line}\approx 1.1$, but not with the relation
(\ref{H=2/D_M}) to $\alpha_{\rm line}\approx 0.74$ or
the mass fractal dimension $D_M\approx 2.7$.
In other words, for the line on a crumpled paper, the self-similarity
of each line structure in the short length scale is not consistent
with the overall scaling upon changing the size $L$ in contrast to the
case of the whole structure of a crumpled paper, in which case
$D_M=d_f$, thus the self-similarity of the internal structure is
consistent with the global scaling.
  The existence of these two regimes for the line structure may 
come from the layered structure of crumpled sheets;
The whole structure of crumpled sheets consists of random folding of
a wrinkled sheet, and typical scale of random folding is of order of
$R_g$ while wrinkling gives shorter length scale with $H\approx 0.9$.

Before concluding, let us discuss some of the previous works using CT
technique.
Lin {\it et al.} \cite{Lin-2009-PRE,Lin-2009-PRL} 
examined the structure of crumpled aluminium foils by CT.
The aluminium foils of different radius $R_0=3\sim 10$ mm are crumpled
into the ball with the same final radius $R=1.5$ mm, i.e. different
compaction ratios.
They estimate the fractal dimension by the box counting and the
correlation dimension as a function of the compaction ratio
$R/R_0$\cite{Lin-2009-PRE}.  Their estimates of the dimensions
coincide fairly well with our estimates for the aluminium foil by the
box counting and the Fourier transform.  They also measured the
correlation for the tangent vector and observed the layered
structure\cite{Lin-2009-PRL}.

Cambou and Menon\cite{Cambou-2011} also used CT to 
examine the internal structure of crumpled aluminium foils.  They
obtained the mass distribution, the distribution of the normal vector
and the curvature radii, and found that they are distributed quite
uniformly.  They also found the layered structure, but again their
orientation is distributed uniformly.  
These uniform distributions might appear to contradict the fractal
structure that has been found in the present work as well as earlier
studies\cite{Balankin-2007b,Lin-2009-PRE,Balankin-2010,Balankin-2013a}.  
They are, however, not contradicting because what they
studied are averaged distributions of the quantities;  Spatial
inhomogeneity of fractal structure varies from a sample to another, 
thus does not likely show in the averaged distribution.

In the present work, we did not examine the scaling relation with the
applied force, Eq.(\ref{R-F_scaling}).
Its exponent $\delta$ should represent how the crumpling energy 
increases as a paper sheet is crumpled into a smaller ball,
thus should come from the self-similarity of the internal structure,
although we do not know yet how it is related with other exponents.

\acknowledgments  This work is partially supported by JSPS KAKENHI
Grant Number JP20K03882.

\vskip 1cm

\bibliography{../references}

\def\thefigure{S\arabic{figure}}
\def\thetable{S\arabic{table}}

\topmargin=-2.5cm \textheight=25cm
\oddsidemargin=0.0cm \textwidth=16cm

\renewcommand{\baselinestretch}{1.1}

\newcolumntype{C}[1]{>{\hfil}m{#1}<{\hfil}}

\setcounter{table}{0}
\setcounter{figure}{0}
\newpage
\vspace*{3cm}

\begin{center} \large\bf

  Supplemental Material to \\
``Fold analysis of crumpled sheet using micro computed tomography''
\vskip 3ex

\normalsize\rm

Yumino Hayase, \;
Hitoshi Aonuma, \; 
Satoshi Takahara, \\
Takahiro Sakaue,  \;
Shun'ichi Kaneko,  \;
and \; Hiizu Nakanishi 
\vskip 2ex

  (\today)
\end{center}
\vskip 4ex

\begin{quote}
  Relaxation effects on paper samples after crumpling
  are examined for some of the scaling exponents that we report
  in the main text.
  It turns out that the effects are negligible already at one hour after
  crumpling in comparison with the overall accuracy of the measurement.
\end{quote}
\vskip 4ex


As being described in the text, the crumpled paper and cellophane
balls are left for 7 days after crumpling before being CT-scanned in
order to allow them to settle.  Some may wonder if  the 7-day waiting
time is enough because it has been observed that crumpled paper balls
undergo slow relaxation logarithmic in time over more than three
weeks.
In order to examine how our results may or may not depend on the
waiting time, we measure the time evolution of some of the quantities
we report in the main text.
\vskip 2ex

The quantities we examined in this supplemental material are
(i) the radius of gyration $R_g$,  (ii) the power spectrum
of the density distribution $S(q)$, and  (iii) the box counting data for the
fractal dimension.  
Five sheets of square paper with the size $L=200$ mm are crumpled by
hands, and are scanned by micro-CT at one hour, 3 days, 7 days, and 14
days after being crumpled.  
The samples are kept under fairly constant condition controlled at the
temperature $24\pm 1^\circ$ C and the humidity $40\pm10\%$.
\vskip 2ex

\newpage
\begin{figure}
  \centerline{\includegraphics[width=18cm]{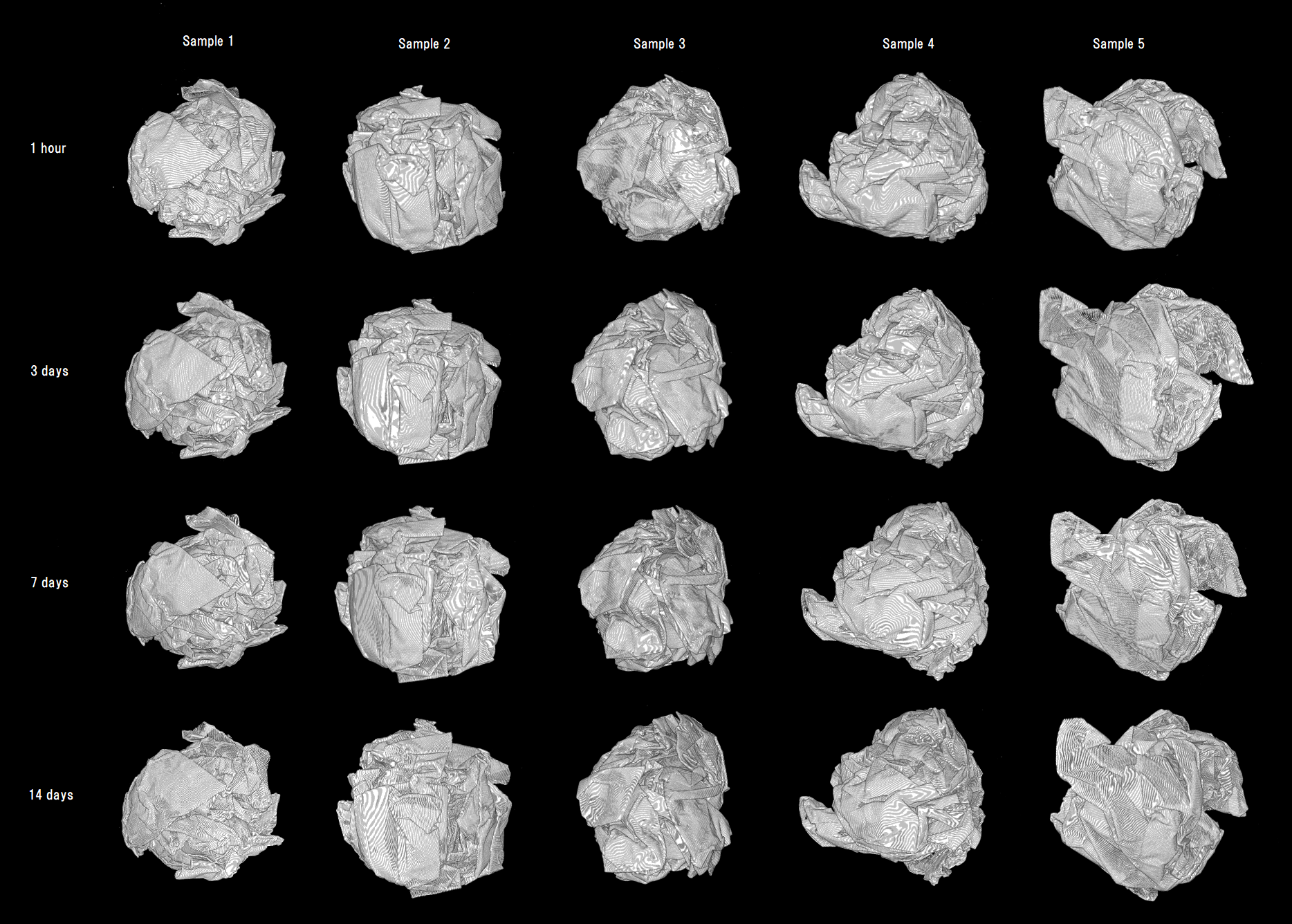}}
  \caption{
  Reconstructed images of five crumpled balls at one hour, 3 days, 7 days,
  and 14 days after being crumpled.}
  \label{crumpled-balls}
\end{figure}
Before we present the measured quantities, we show the reconstructed
images of the samples upon each scan in Fig.\ref{crumpled-balls}.  The
sample orientation is not controlled at the scanning, thus the lines
of sight for these images are adjusted on the computer by 3d viewer
after each image is reconstructed.  One may notice some relaxation of
the configuration, but they do not change much after one hour.
\vskip 2ex

\newpage
The calculated radii of gyration are listed in Table \ref{table-SI}.
They increase slightly over time, but
the change in the radius for each sample after one hour is less than 1
mm;  it is smaller than variation among samples, and should not have
any effect on the estimate of the exponent $\alpha\approx 0.74$ in
Fig.3(b) of the main text.
The last data of Sample 5 is smaller than the earlier ones.  We do not
understand how this happened, but
a certain relaxation in the density inhomogeneity
inside the ball could yield such a fluctuation of $R_g$.
\vskip 2ex

\begin{table}[t]
  \centerline{
  \begin{tabular}{c|ccccc}
  time & Sample 1 & Sample 2 & Sample 3 & Sample 4 & Sample 5 \\
    \hline\hline
1 hour & 12.31    & 11.48    & 11.58    & 12.04    & 12.44 \\
3 days & 12.52    & 11.52    & 11.67    & 12.16    & 12.71 \\
7 days & 12.83    & 11.76    & 11.99    & 12.45    & 13.21 \\
14 days & 13.11   & 11.97    & 12.28    & 12.85    & 12.49 \\
  \end{tabular}
}
  \caption{
    The radii of gyration $R_g$ [mm] of each sample at 1 hour, 3 days, 
    7 days, and 14 days after being crumpled.
    They are calculated from binarized data of the images.
}
  \label{table-SI}
\end{table}

\begin{figure}[t]
  \centerline{\includegraphics[width=13cm]{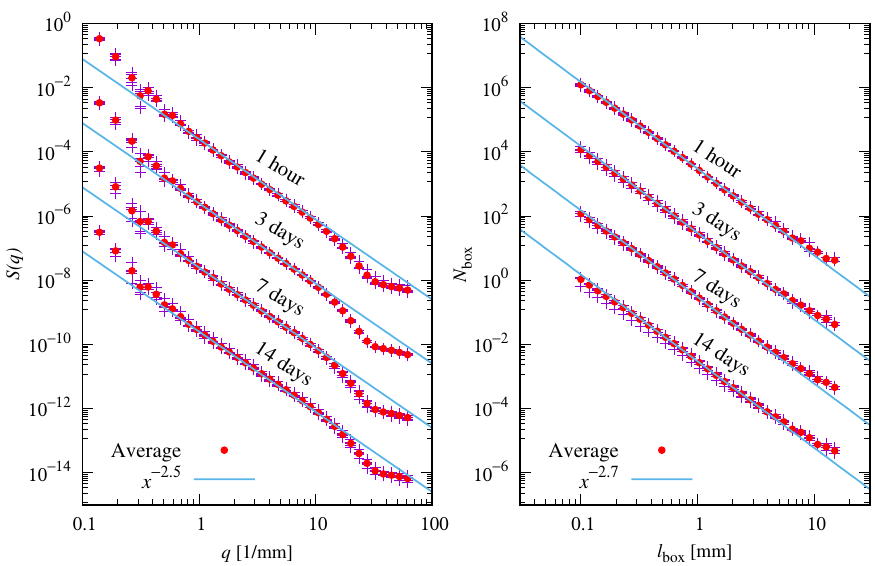}}
  \caption{The evolution of the power spectra $S(q)$ and the box
  counting data $N_{\rm box}$ v.s. $l_{\rm box}$.
  The data for different times are shifted by the factor $10^{-2}$.}
  \label{exponent}
\end{figure}

Figure \ref{exponent} shows the evolution of the power spectra $S(q)$
of the density distributions and the box counting data $N_{\rm box}$
v.s. $l_{\rm box}$ for the fractal dimension; the plus marks show the
data for each sample and the red solid circles show the average over
the 5 samples.  The lines are the scaling fits by $S(q)\sim
q^{-\beta}$ and $N_{\rm box}\sim l_{\rm box}^{-d_f}$ with $\beta=2.5$
and $d_f=2.7$.  The data for different scan times are shifted by the
factor $10^{-2}$.  
One may find small changes over time at the both ends of $q$ or $l_{\rm
box}$, but the data 
are quite stable
in the middle range, where the exponents are estimated, therefore,  
the estimates for the exponent $\beta$ and $d_f$ do not depend on the
waiting time
for all the data we examine.
\vskip 2ex


\end{document}